\documentclass[fleqn,usenatbib]{mnras}

\usepackage{newtxtext,newtxmath}

\usepackage{graphicx,subfig}	
\usepackage[T1]{fontenc}
\usepackage{ae,aecompl}

\usepackage{graphicx}	
\usepackage{amsmath}	
\usepackage{amssymb}	

\title[Spectral performance of Square Kilometre Array Antennas II: Calibration performance]{Spectral performance of Square Kilometre Array Antennas II: Calibration performance}

\author[C. M. Trott]{Cathryn M. Trott$^{1,2}$\thanks{E-mail: cathryn.trott@curtin.edu.au},
Eloy de Lera Acedo$^3$,
Randall B. Wayth$^{1,2}$,
Nicolas Fagnoni$^{3}$,
\newauthor
Adrian T. Sutinjo$^{1}$,
Brett Wakley$^{4}$,
Chris Ivan B. Punzalan$^{1}$
\\
$^{1}$International Centre for Radio Astronomy Research (ICRAR), Curtin University, Bentley Australia\\
$^{2}$Australian Research Council Centre of Excellence for All-Sky Astrophysics (CAASTRO), Australia\\
$^{3}$Cavendish Laboratory, University of Cambridge, Cambridge, CB3 0HE, United Kingdom\\
${^4}$Cambridge Consultants Ltd, Cambridge, United Kingdom\\
}

\pubyear{2017}

\begin{document}
\label{firstpage}
\pagerange{\pageref{firstpage}--\pageref{lastpage}}
\maketitle

\begin{abstract}
We test the bandpass smoothness performance of two prototype Square Kilometre Array (SKA) SKA1-Low log-periodic dipole antennas, the SKALA2 and SKALA3 (`SKA Log-periodic Antenna'), and the current dipole from the Murchison Widefield Array (MWA) precursor telescope. Throughout this paper, we refer to the output complex-valued voltage response of an antenna when connected to a low noise amplifier (LNA), as the dipole bandpass.
In Paper I \citep{deleraacedo16}, the bandpass spectral response of the log-periodic antenna being developed for the SKA1-Low was estimated using numerical electromagnetic simulations and analyzed using low-order polynomial fittings and it was compared with the HERA antenna against the delay spectrum metric.
In this work, realistic simulations of the SKA1-Low instrument, including frequency-dependent primary beam shapes and array configuration, are used with a weighted least-squares polynomial estimator to assess the ability of a given prototype antenna to perform the SKA Epoch of Reionisation (EoR) statistical experiments.
This work complements the ideal estimator tolerances computed for the proposed EoR science experiments in \citet{trottwayth2016}, with the realised performance of an optimal and standard estimation (calibration) procedure.
With a sufficient sky calibration model at higher frequencies, all antennas have bandpasses that are sufficiently smooth to meet the tolerances described in \citet{trottwayth2016} to perform the EoR statistical experiments, and these are primarily limited by an adequate sky calibration model, and the thermal noise level in the calibration data.
At frequencies of the Cosmic Dawn (CD), which is of principal interest to SKA as one of the first next-generation telescopes capable of accessing higher redshifts, the MWA dipole and SKALA3 antenna have adequate performance, while the SKALA2 design will impede the ability to explore this era.
\end{abstract}




\section{Introduction}
The Square Kilometre Array \citep[SKA]{dewdney16} will be the largest radio telescope in the world, upon construction in the coming years. The low-frequency component, SKA1-Low, operating between 50 and 350~MHz, will be sited in the Western Australian desert, with its core within the Murchison Radio astronomy Observatory. Its primary science drivers include detection and timing of southern hemisphere pulsars, detection and characterisation of Fast Radio Bursts (FRBs), deep continuum and neutral hydrogen surveys, and detection and exploration of the Cosmic Dawn ($z=15-25$) and Epoch of Reionisation ($z=5.5-15$). SKA aims to be the first telescope to perform direct spectral line imaging of  structures at high redshift (21cm tomography), beyond the statistical experiments of current low-frequency radio telescopes \citep{koopmans15}.

The weakness of the EoR/CD signal relative to system noise (particularly at low frequency, where sky noise dominates the antenna temperature), and foreground continuum sources (Galactic and extragalactic), and the intrinsic chromaticity of aperture array interferometers, yield a challenging experiment. Current low-frequency telescopes that are attempting to detect the EoR signal at lower redshift (e.g., Murchison Widefield Array (MWA) \citep{lonsdale09,tingay13_mwasystem,jacobs16}; Precision Array for Probing the Epoch of Reionization (PAPER) \citep{parsons10}; the Low Frequency Array (LOFAR) \citep{vanhaarlem13,patil16}; the Long Wavelength Array (LWA) \citep{ellingson09}) have struggled with instrument calibration and foreground removal, and a detection has thus far eluded all. The additional sensitivity of SKA, and the key design features of HERA \citep[Hydrogen Epoch of Reionization Array]{deboer16}, improve the prospects for EoR science, but substantial challenges remain.

An important aspect of the system design is the response of the receiving antenna as a function of frequency (spectral response) to incoming signals. 
Recent experience with the MWA has highlighted how uncalibrated (or uncalibratable) structure in the antenna bandpass can corrupt the EoR power spectrum measurements.
Work by \citet{barry16} to simulate the effects of antenna cable reflections in the MWA signal chain demonstrated the requirement for smoothness across the bandpass.
More recently, \citet{offringa16} and \citet{ewallwice16} studied the impact of calibration errors due to unmodelled foreground sources, and found that residual source chromatic structure impedes the ability of instruments to perform EoR science.
Relevant to this work, \citet{trottwayth2016} used an ideal estimator to quantify the tolerances on smoothness of the intrinsic SKA-Low bandpass response required for the proposed EoR/CD experiments to be undertaken successfully.
To test the tolerances of this ideal estimator analysis, in this work we perform realistic simulations of data and calibration using simulated measurements of the spectral response of the SKALA prototype, and actual measurements of the spectral response of the MWA Engineering Development Array (EDA) antennas, to demonstrate their capacity to meet the required tolerances.

The antenna response simulations described in Paper I~\citep{deleraacedo16} demonstrate the ability of the new SKALA3 design to meet the tolerances described in \citet{trottwayth2016} and a performance comparable to that of the HERA dish antenna~\citep{deboer16} when measured against the delay spectrum metric. However, that work computed estimation bounds based on a theoretical ideal estimator (efficient and unbiased), using the Fisher Information. In general, this optimal estimator may not exist.
The residuals of this optimal estimation procedure were then propagated to the EoR power spectrum, and compared with the expected thermal noise. These propagated errors defined the tolerances by being forced to lie at lower amplitude than the experimental thermal noise. In addition to the requirement for an ideal estimator, the simulated measurements of the antenna response only compare the performance of one component of a complicated signal chain, and coupling between other components may reduce calibration performance.

To address these concerns, here we perform realistic simulations at 65, 100, 150, and 200~MHz of SKA1-Low. To the instrument, we employ models for the point source population of the sky (using measured properties of the spatial and flux density distribution of extragalactic radio sources) and apply an optimal and standard calibration procedure to the simulated data. This procedure entails using a weighted least-squares estimator to fit a function across the spectral channels of simulated interferometer measurements, relative to an expected measurement from a model of the signal in the sky. This procedure uses a $n$-th order polynomial fit ($n=1-4$) of the real and imaginary components of a station bandpass, across 504 5~kHz fine channels (three coarse channels), and uses these parameters to calibrate the central coarse channel, using a weighted least squares estimator. This directly compares to one of the general calibration approaches described in \citet{trottwayth2016}. 

\section{Antennas}
In Paper I, the design and performance of the SKALA2 and SKALA3 are presented, and we summarise those here. The bandpass here is the complex-valued voltage gain of the Low Noise Amplifier (LNA)
when connected to the antenna impedance, as a function of frequency. It is simulated for a single dipole system. The SKALA2 antenna is a log-periodic dipole array made of four identical metallic arms. These form two polarizations, and yield a design with large fractional bandwidth and high-directivity. The SKALA2 footprint is 1.2~m wide at the base, and it is 1.8~m tall. The SKALA3 has the same basic structural design, but has been modified to alleviate features in the spectral response due to impedance mismatch between the antenna and low-noise amplifier. To allow this, the bottom dipole of the antenna has been enlarged, yielding a broader footprint of 1.6~m in breadth.

We use the simulated voltage output of each SKALA antenna as a function of frequency as the reference bandpass in this work. We further use actual measurements of the MWA EDA dipole for its reference bandpass.
The SKALA antenna designs, electromagnetic simulations, and output spectral bandpasses, including the effects of the LNAs connected to the antenna, are described in \citet{deleraacedo16} (Paper I), and we use those responses in this work. As a reference to an existing precursor telescope, we compare the SKALA2 and SKALA3 performance with that of the dipole antenna of the MWA.

The SKALA antennas are log-periodic dipoles, designed to have good sensitivity across a wide fractional bandwidth, to undertake a range of science experiments~\citep{deLeraAcedo2015}. In particular, one of the key design drivers of SKALA is the maximization of its sensitivity across the band 50 to 350 MHz and even up to higher frequencies. Sensitivity in this context is measured as the effective area over the system temperature. The system temperature, while dominated by the sky noise at low frequencies, is heavily dominated by the receiver noise in the upper half of the SKA1-Low band. This maximization calls for a delicate trade off between the power match to the LNA (for a smooth passband response) and the noise match to the LNA (for minimum noise generated by the receiver) as explained in~\citet{deleraacedo16}. There are other fundamental antenna designs, with advantages and disadvantages compared with the log periodic design.

We use a dipole from the EDA, which is a test system composed of 256 MWA dipoles in a random configuration, with a station diameter of 35\,m \citep{wayth16}.
The dipoles and LNAs used in the EDA are identical to the MWA's dipoles with the exception that the lowest frequency of
the band has been reduced from 70 MHz (as for the MWA) to approximately 50 MHz, by changing a passive component value.

Unlike the SKALA antennas, which have been simulated electromagnetically, the EDA bandpass is based on measurement of a single MWA dipole antenna over a $1.81\times 1.81$\,m$^{2}$ ground mesh using an EDA LNA. We expect the measurement to offer a more realistic characterisation of the bandpass at the expense of some measurement noise.

\begin{figure*}
\subfloat[SKALA2 bandpass amplitude (scaled) and phase.]{
\includegraphics[width=.32\textwidth]{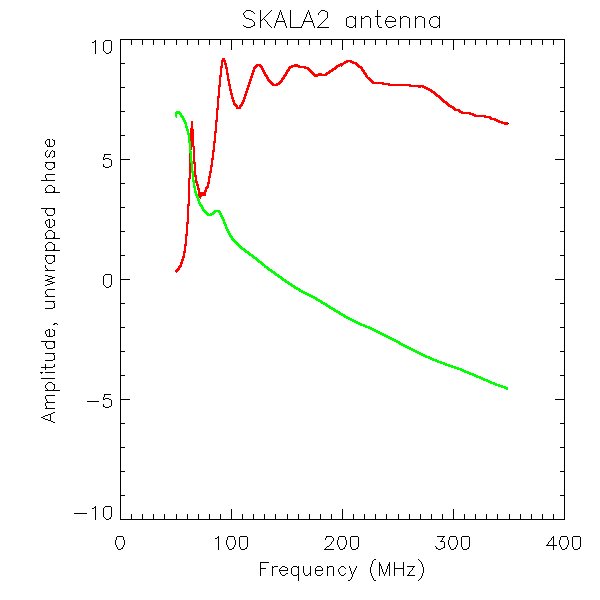}
}
\subfloat[SKALA3 bandpass amplitude (scaled) and phase.]{
\includegraphics[width=.32\textwidth]{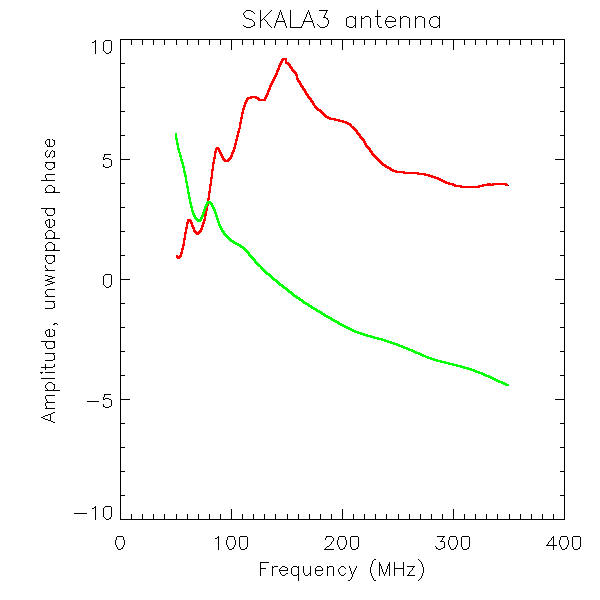}
}
\subfloat[MWA EDA bandpass amplitude (scaled) and phase.]{
\includegraphics[width=.32\textwidth]{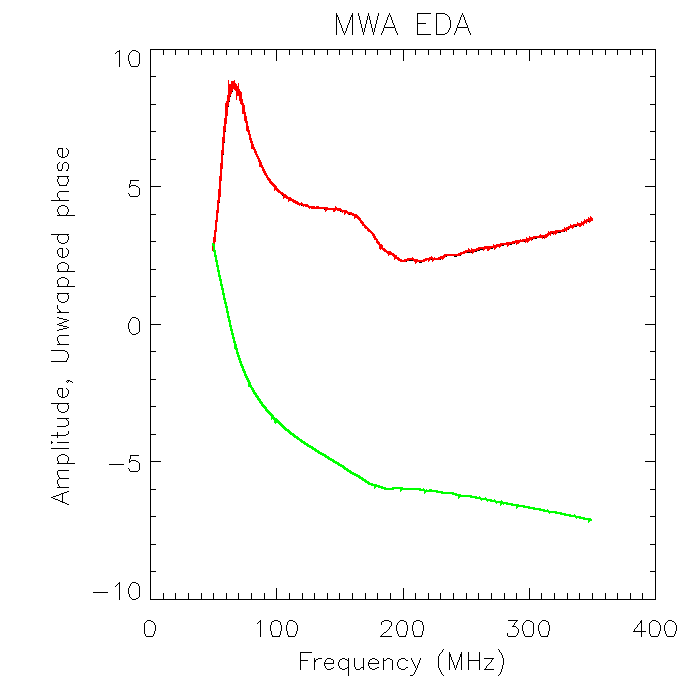}
}
\caption{Left: Simulated bandpass amplitude (red) and unwrapped phase (green, radians) of the SKALA2 antenna-LNA system (linear units). Centre: same, but for SKALA3. Right: MWA EDA. The amplitudes are scaled to have the same axis units as the phase.}
\label{fig:skala_eda_bp}
\end{figure*}

A reproduction of the simulated bandpass of the SKALA antenna/LNA system is shown in Figure \ref{fig:skala_eda_bp} (left, centre), where the amplitude has been scaled for ease of plotting with the phase (radians). I.e., amplitude and phase have the same axis scale. The corresponding measured response for the MWA dipole is also shown (right).
Note the reduced level of sharp amplitude peaks in the SKALA3, compared with the SKALA2, particularly at low frequencies.

\begin{figure}
\includegraphics[width=.45\textwidth]{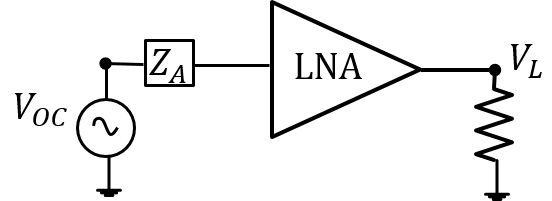}
\caption{Diagrammatic view of the measurement of the EDA bandpass, where the response is taken to mean the direction independent voltage ratio of the voltage across the load at the output of the LNA to the antenna open circuit voltage. The load resistance is $50\,\Omega$.
In reality, the bandpass response includes the entire signal chain from the antenna open circuit voltage to the voltage at the input of the correlator. Our present scope, however, is in understanding the interaction between the antenna impedance and the LNA input impedance particularly at low frequencies where the antenna becomes electrically small causing significant mismatch, which may result in peaking response in the passband. The diagram may be extended to include more components in the signal chain by cascading their two-port representations.}
\label{fig:passband}
\end{figure}

In the context of this work, we take the MWA bandpass response to mean the \emph{direction independent voltage ratio of the voltage across the load at the output of the LNA to the antenna open circuit voltage.}
This is depicted in Fig. \ref{fig:passband} as the $V_{L}/V_{OC}$.
In the case of the MWA dipole, the antenna impedance, $Z_{A}$, is the measured differential impedance. The two-port LNA parameters are obtained through vector network analyzer (VNA) measurements. The measurements have 10\,kHz spectral resolution between $\nu=$\,50 -- 350\,MHz. 


\section{Simulations}
The simulations include the current SKA1-Low array configuration, realistic stations formed by randomly distributing 256 dipoles over a diameter of 35\,m (with minimum spacing 1.5\,m), and a realistic sky model. The simulated visibilities are attenuated by the antenna bandpass, and noise is added. The noise is uncorrelated between stations, with a frequency-dependent system temperature, as described in \citet{turner16}.
The data are then calibrated using the approach described, and residuals are computed as a function of frequency.
These residuals are directly compared with the tolerances described in \citet{trottwayth2016}.
These are reproduced in Table \ref{table:tolerances} for reference.

\begin{table}
\centering
\begin{tabular}{|l|l|l|l|}
\hline 
Experiment & $\delta_{n=2}$ & $\delta_{n=3}$ & $\delta_{n=4}$\\
\hline
50~MHz & 0.027 & 0.025 & 0.019\\
100~MHz & 0.011 & 0.010 & 0.008\\
150~MHz & 0.006 & 0.005 & 0.004\\
200~MHz & 0.009 & 0.008 & 0.006\\
\hline
\end{tabular} 
\caption{Derived tolerances for each experiment such that the residual power due to ($n$+1)th-order curvature in the bandpass is less than the thermal noise. Reproduced from \citet{trottwayth2016}.}
\label{table:tolerances}
\end{table}

The approach to testing the calibration performance of the array is:
\begin{enumerate}
\item We define 504 5\,kHz fine channels (2.52\,MHz bandwidth), corresponding to three 840\,kHz coarse channels of 168 fine channels each, at a given reference frequency in the 60--250\,MHz band;
\item The zenith $uv$-distribution of the array is created using the V4A design (SKAO Science Team, SKA-SCI-LOW-001, 2015);
\item A sky model point source population is simulated, with $S_{\rm Jy}=0.03-100.0$\,Jy, according to a number density model for bright sources, which is composed of a broken power law, with break at 1\,Jy:
\begin{eqnarray}
\langle{N(S,S+dS)}\rangle(\nu) &=& \frac{dN}{dS}(\nu)\,dS\,d{\bf l} \\
&=& \alpha \left( \frac{\nu}{\nu_0} \right)^{\gamma} \left( \frac{S_{\rm Jy}}{S_0}\right)^{-\beta}\,dS\,d{\bf l}.
\label{source_counts}
\end{eqnarray}
We use values of $\alpha=3900\,{\rm Jy}^{-1}{\mathrm sr}^{-1}$, $\beta_{>\rm{brk}}=2.50$, $\beta_{<\rm{brk}}=1.59$ and $\gamma=-0.8$ at 150~MHz, in line with recent measurements for the high flux density part of the distribution \citep{intema11,franzen16,williams16}. Sources are distributed uniformly in flux density bins, and across the sky;
\item A frequency-dependent station beam is formed by evenly-distributing 256 dipole antennas within a 35\,m diameter circle, and perturbing their positions.
This layout is used to compute the beam for each fine channel;
\item The sky model is attenuated by the beam at each frequency, and sources above a threshold are retained in the calibration model;
\item For each fine channel, calibration sources are gridded onto the sky, with spatial resolution exceeding the Nyquist sampling for the longest baseline in the array in the $uv$-plane;
\item For each fine channel, the sky model is Fourier Transformed to the $uv$-plane, and visibilities are sampled according to a nearest-neighbour sampling;
\item Antenna bandpass measurements (simulated for the SKALA antennas, measured for the MWA dipole) are applied to the real and imaginary components of the visibilities, and thermal noise is added to mimic the expected level for each fine channel with a calibration cycle of 600\,s;
\item The calibration proceeds in two steps:
$(1)$ form the initial calibration estimate, by taking the ratio of the measured visibilities to the expected, given the calibration sky model, and the weighted-mean for each station is computed across all baselines in which it participates;
$(2)$ These initial parameter fits are then used as initial estimates in a Levenberg-Marquardt minimisation of polynomial parameters.  An nth-order polynomial is fitted across the band for a given station bandpass, using a least-squares estimator weighted by the channel noise (IDL poly\_fit procedure). Residuals are computed by subtracting the actual noise-free visibilities from the fitted.

\end{enumerate}

There are three sky models used in the calibration: `Perfect', `Imperfect(1)', and `Imperfect(2)':
\begin{itemize}
\item Perfect: uses the same model as was generated to simulate the visibilities, and therefore represents the best, unbiased data model;
\item Imperfect$(1)$: uses the brightest sky sources, but sources below a high calibration threshold (both resolved and confused) remain in the data, yielding a noise-like, unmodelled signal: $S_{\rm thresh}=0.30$\,Jy;
\item Imperfect$(2)$: uses the brightest sky sources, but sources below a low calibration threshold (both resolved and confused) remain in the data, yielding a noise-like, unmodelled signal: $S_{\rm thresh}=0.08$\,Jy.
\end{itemize}

Imperfect(1) is therefore a poorer representation of the sky signal than Imperfect(2). It is expected that the imperfect, but realistic models will yield reduced calibration performance compared with the perfect model.

These simulated visibilities are used to fit for the parameters of a complex-valued station bandpass. The relative residual error, in each fine channel, is defined by the ratio of the residual in a visibility to the actual noise-free visibility (sky model multiplied by the actual bandpass):
\begin{equation}
\epsilon = \left|\frac{\rm{DATA} - \rm{FIT}\times\rm{CAL. MODEL}}{\rm{ANTENNA}\times\rm{SKY MODEL}}\right|,
\end{equation}
where the sky model is complete, the calibration model may be complete or incomplete, and the `ANTENNA' refers to the measured or simulated spectral response of the antenna. Here, `DATA' refers to the simulated real and imaginary components of the visibilities as a function of frequency channel, formed by multiplying the bandpass with the real sky emission.

The fits are performed for the real and imaginary components of the antenna bandpass, and the error in the amplitude formed from these. We average the error, $\epsilon$, over baselines formed with a reference antenna, and over noise realisations. 
Experiments demonstrate that different sky realisations produce comparable results.

\begin{figure}
\hspace{-0.8cm}
{
\includegraphics[width=0.55\textwidth,angle=0]{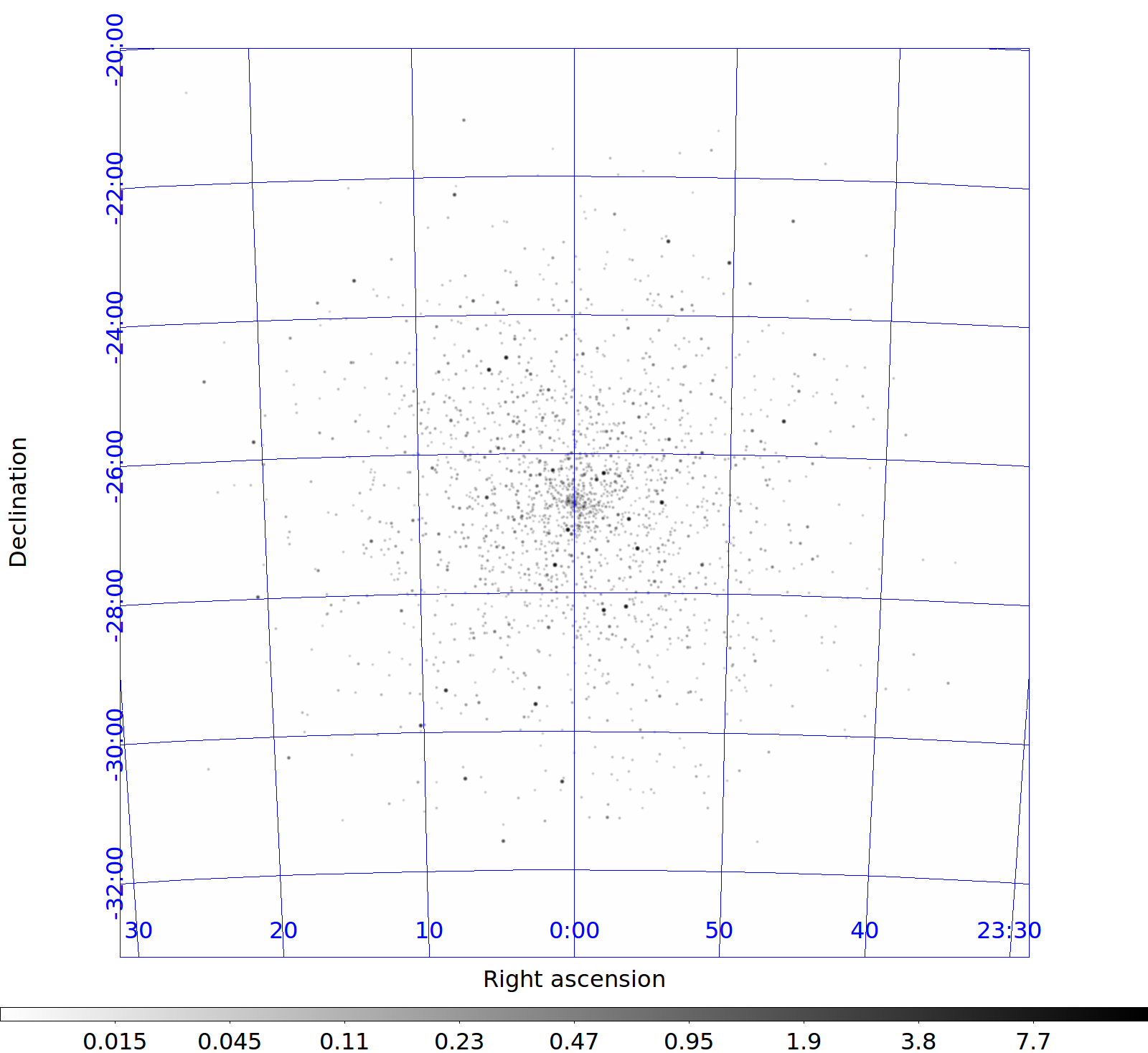}
}
\caption{Example sky source model, weighted by the primary beam of an SKA1-Low station at 150~MHz. Units are Jy/beam, with 12 Jy/beam as the brightest source in the field.}
\label{fig:sky}
\end{figure}

Figure \ref{fig:sky} displays an example sky model, attenuated by the primary beam.

\section{Results}
We begin by presenting a comparison of the SKALA2 and SKALA3 calibration performance, because these are directly comparable and pertain to the simulations and improvements in design from Paper I. We then include the results for the MWA EDA dipole.

\begin{figure*}
\subfloat[65~MHz.]{
\includegraphics[width=.45\textwidth]{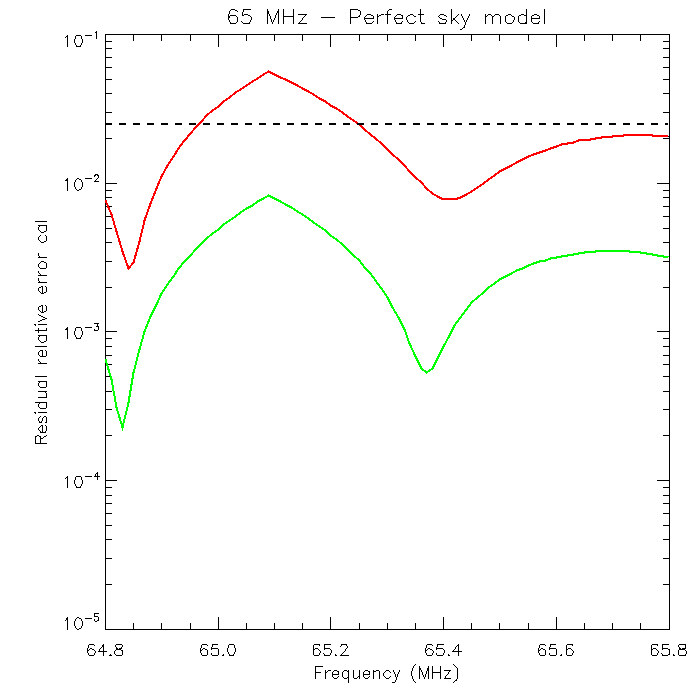}
}
\subfloat[100~MHz.]{
\includegraphics[width=.45\textwidth]{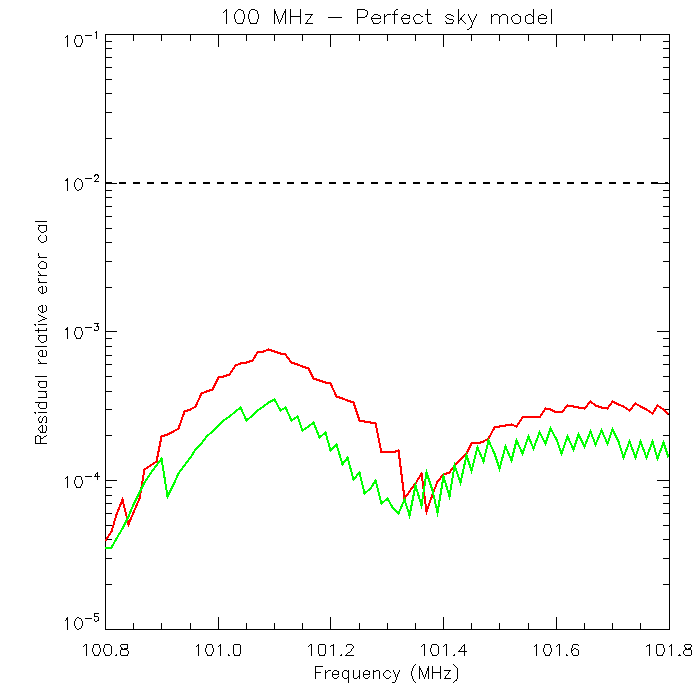}
}\\
\subfloat[150~MHz.]{
\includegraphics[width=.45\textwidth]{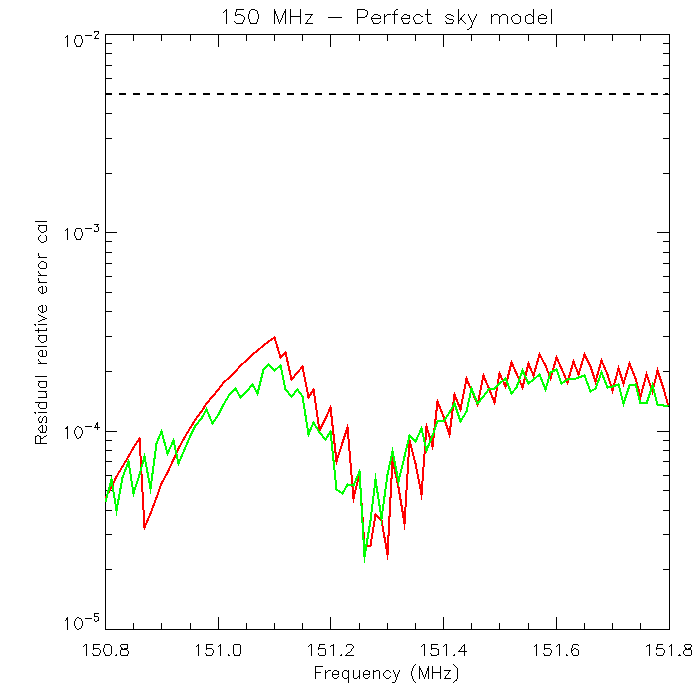}
}
\subfloat[200~MHz.]{
\includegraphics[width=.45\textwidth]{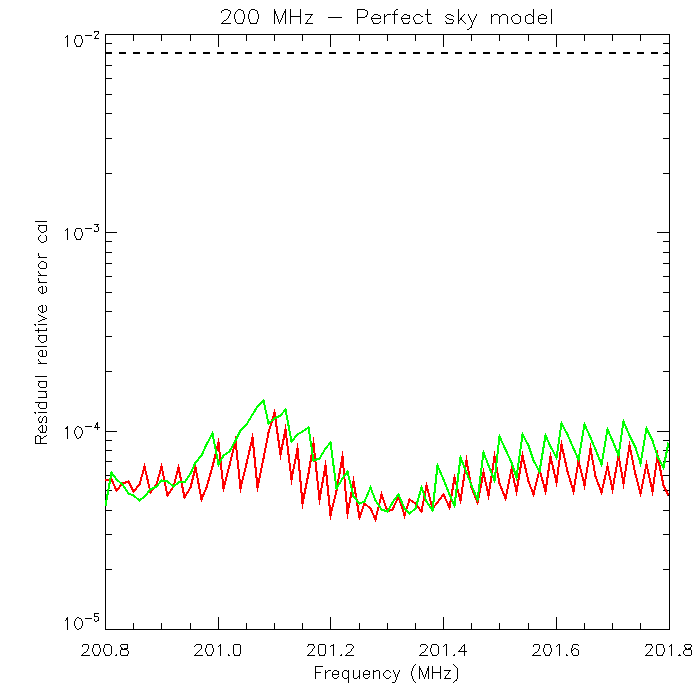}
}
\caption{Residuals in the amplitude of the data, relative to a gain of unity, for four relevant reference frequencies, and the `Perfect' sky model, a third-order polynomial. In all cases, red denotes the SKALA2, and green denotes the SKALA3, design. The dashed, black line denotes the tolerance level proposed in \citet{trottwayth2016}.}
\label{fig:res_sims}
\end{figure*}

\begin{figure*}
\subfloat[65~MHz.]{
\includegraphics[width=.45\textwidth]{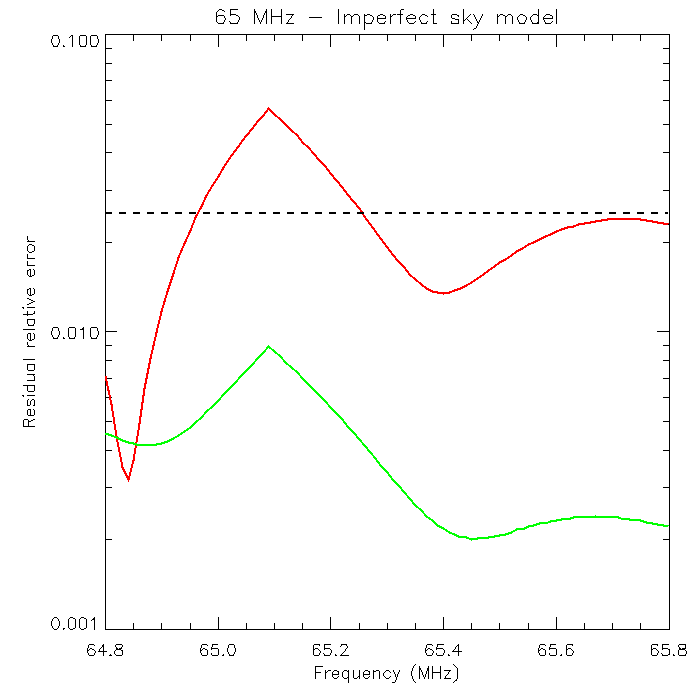}
}
\subfloat[100~MHz.]{
\includegraphics[width=.45\textwidth]{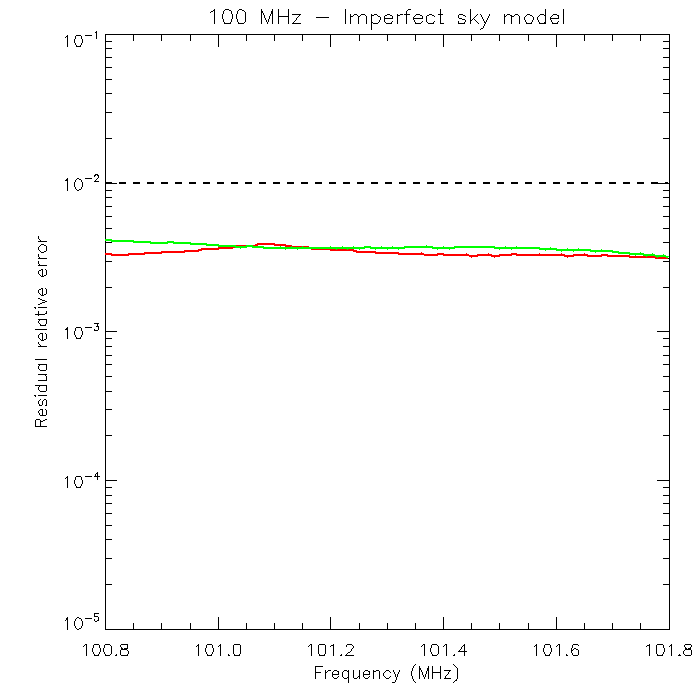}
}\\
\subfloat[150~MHz.]{
\includegraphics[width=.45\textwidth]{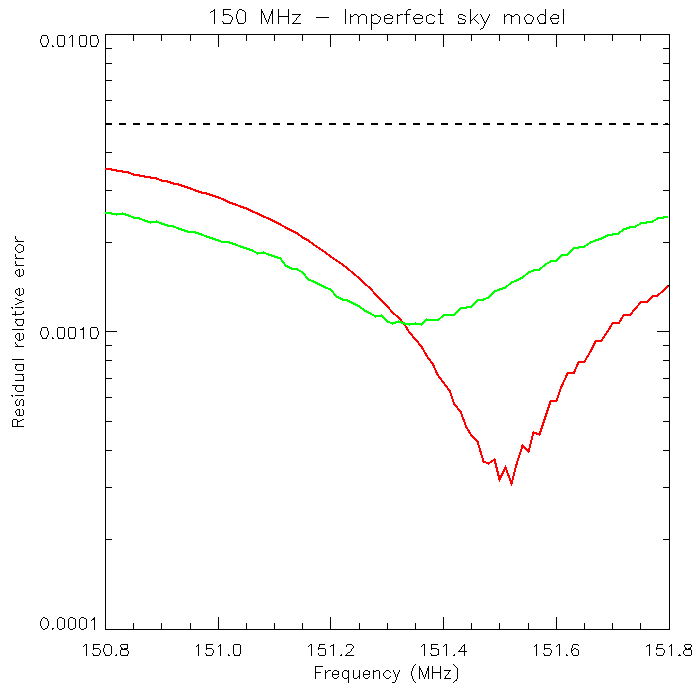}
}
\subfloat[200~MHz.]{
\includegraphics[width=.45\textwidth]{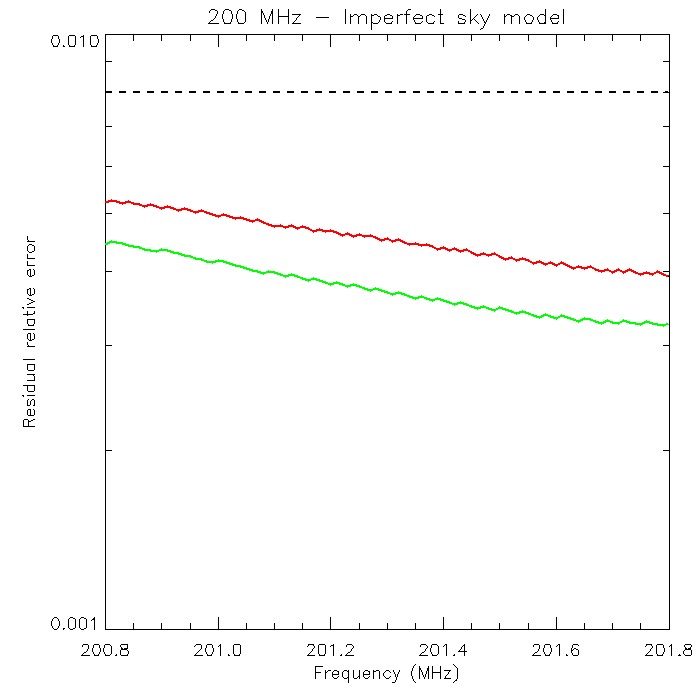}
}
\caption{Residuals in the amplitude of the data, relative to a gain of unity, for four relevant reference frequencies, and an imperfect calibration model (`Imperfect(1)') with only sources with $S>0.3$\,Jy included, a third-order polynomial. In all cases, red denotes the SKALA2, and green denotes the SKALA3, design. The dashed, black line denotes the tolerance level proposed in \citet{trottwayth2016}.}
\label{fig:res_sims_imperfect1}
\end{figure*}

\begin{figure*}
\subfloat[65~MHz.]{
\includegraphics[width=.45\textwidth]{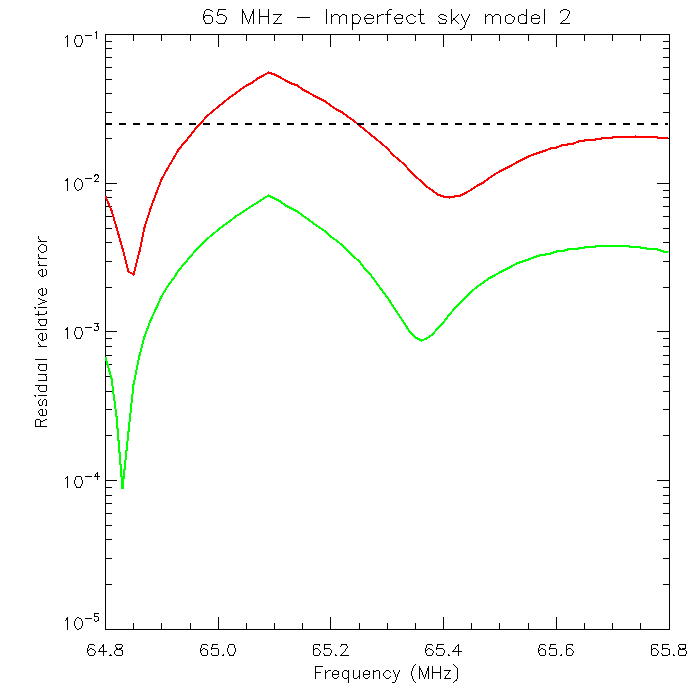}
}
\subfloat[100~MHz.]{
\includegraphics[width=.45\textwidth]{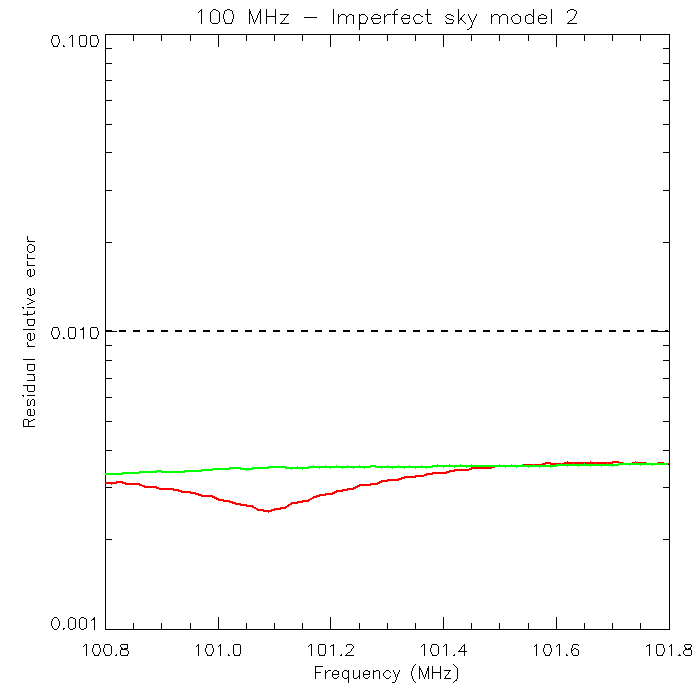}
}\\
\subfloat[150~MHz.]{
\includegraphics[width=.45\textwidth]{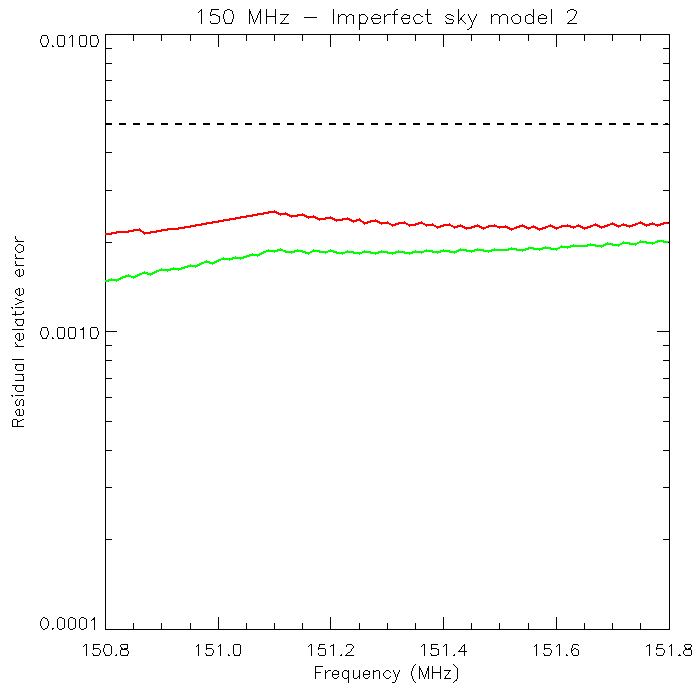}
}
\subfloat[200~MHz.]{
\includegraphics[width=.45\textwidth]{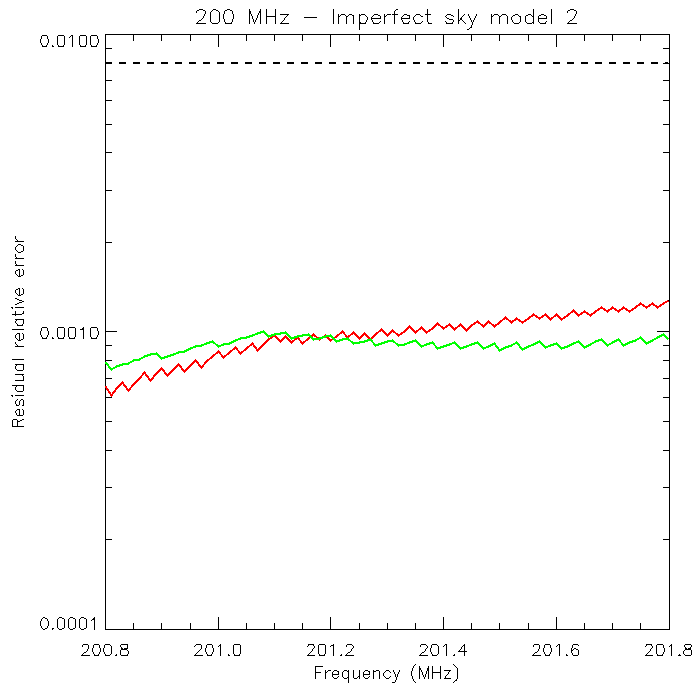}
}
\caption{Residuals in the amplitude of the data, relative to a gain of unity, for four relevant reference frequencies, and an imperfect calibration model (`Imperfect(2)') with only sources with $S>80$~mJy included, a third-order polynomial. In all cases, red denotes the SKALA2, and green denotes the SKALA3, design. The dashed, black line denotes the tolerance level proposed in \citet{trottwayth2016}.}
\label{fig:res_sims_imperfect2}
\end{figure*}

\begin{figure}
{
\includegraphics[width=0.45\textwidth,angle=0]{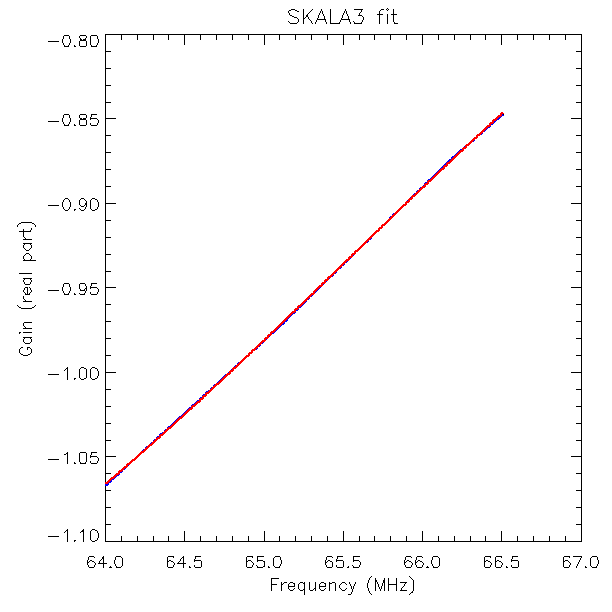}
}
\caption{Real-part of the bandpass (blue) and polynomial fit (red) for the SKALA3 antenna at 65~MHz.}
\label{fig:fit}
\end{figure}

Figure \ref{fig:res_sims} displays residuals in the amplitude of the data, relative to a gain of unity, for four relevant reference frequencies, where the error is computed for the central coarse channel, and the `Perfect' calibration model. The two imperfect calibration models (`Imperfect(1)' and `Imperfect(2)') are displayed in Figures \ref{fig:res_sims_imperfect1} and \ref{fig:res_sims_imperfect2}, respectively.
An example of the real part of the SKALA3 simulated model and fitted polynomial is shown in Figure \ref{fig:fit}.


Inspection of Figures \ref{fig:res_sims}, \ref{fig:res_sims_imperfect1} and \ref{fig:res_sims_imperfect2} demonstrate the following:
\begin{itemize}
\item The SKALA2 bandpass error is always above the threshold for all calibration models around 65\,MHz;
\item Residuals for the calibration error tend to go down with improved sky model, as expected.
\end{itemize}

The consistency of the result at the lowest frequency is demonstrating that a 600~s calibration cycle is more thermal-noise dominated (rather than sky-model error dominated), and therefore noise plays a major role in the relative antenna performance.

For the poorest calibration model, the SKALA2 performance is degraded, and for all cases the SKALA2 yields poor performance compared with the SKALA3. At higher frequencies, the thermal noise level is relatively lower, and the calibration model errors (residual noise-like signal from other, unmodelled sources in the sky) are more dominant in the fitting error budget. These results are generally true, but the transition from sky- to noise-dominated will depend on frequency, calibration cycle time, and sky model accuracy.

The second notable feature is the relatively-similar performance of the two antennas for $\nu>$~100~MHz. Here, the error in the fitting, and therefore the residual, has increased, due to the additional unmodelled sky signal in the data, and the intrinsic spectral performance of the antennas has a lesser impact.

When comparing the results with the tolerances determined in \citet{trottwayth2016}, 65~MHz is the only frequency for the SKALA2 antenna where the calibration residuals exceed the tolerances. This is true regardless of the depth of the sky model used for the calibration. 
In both antenna models, the perfect sky model performs the best, and the performance is comparable between designs.

\subsection*{Sky and calibration model}
The sky and calibration model used above are both shallow, given the sensitivity of SKA1-Low. That is, the lower flux density limit of 30~mJy is high, relative to the expected noise level in an image with an integration of 600-seconds (3~mJy at 200~MHz, rising to 15~mJy at 65~MHz, over a 10~kHz band). Therefore, will a deeper sky and/or calibration model improve the performance? To test this, we performed similar simulations with a floor of 5~mJy, thereby including many additional sources in the sky and calibration model, and additional flux. The results were not significantly changed.

\subsection*{Calibration bandwidth}
\begin{figure*}
\subfloat[Perfect calibration model.]{
\includegraphics[width=.45\textwidth]{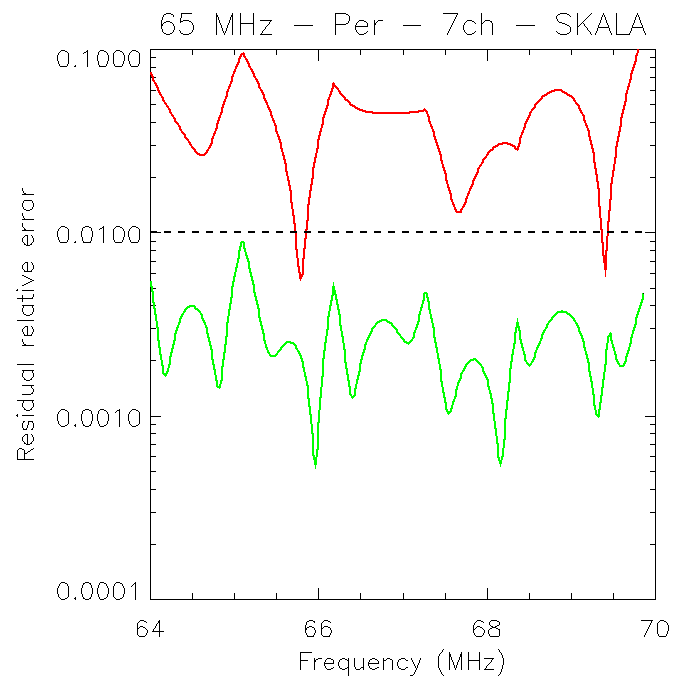}
}
\subfloat[Imperfect 2 calibration model.]{
\includegraphics[width=.45\textwidth]{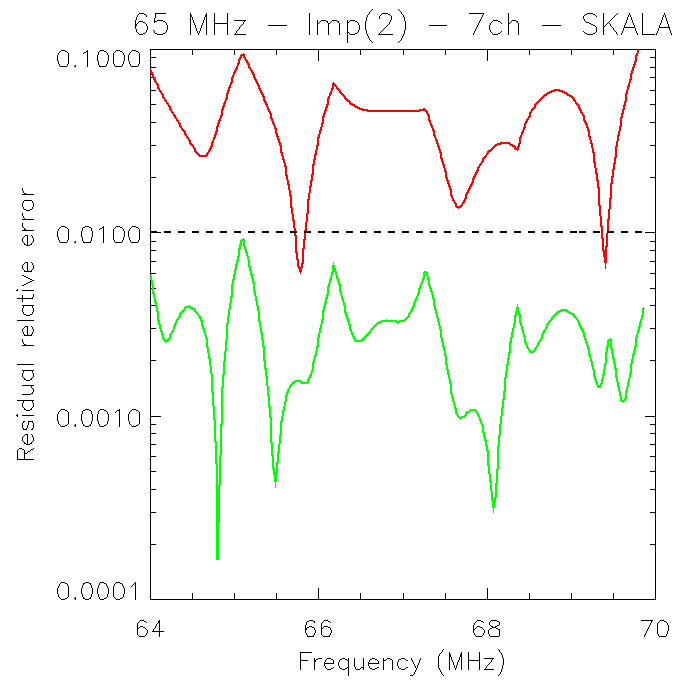}
}
\caption{Relative residual errors for each antenna at 65~MHz, for the perfect and imperfect 2 calibration models, a third-order polynomial, and 7 coarse channels of bandwidth. Also plotted is the tolerance level (black, dashed). The wider fitting bandwidth is reflected in the breadth of the frequency axis shown.}
\label{fig:calibration_bw_7}
\end{figure*}

\begin{figure*}
\subfloat[Perfect calibration model.]{
\includegraphics[width=.45\textwidth]{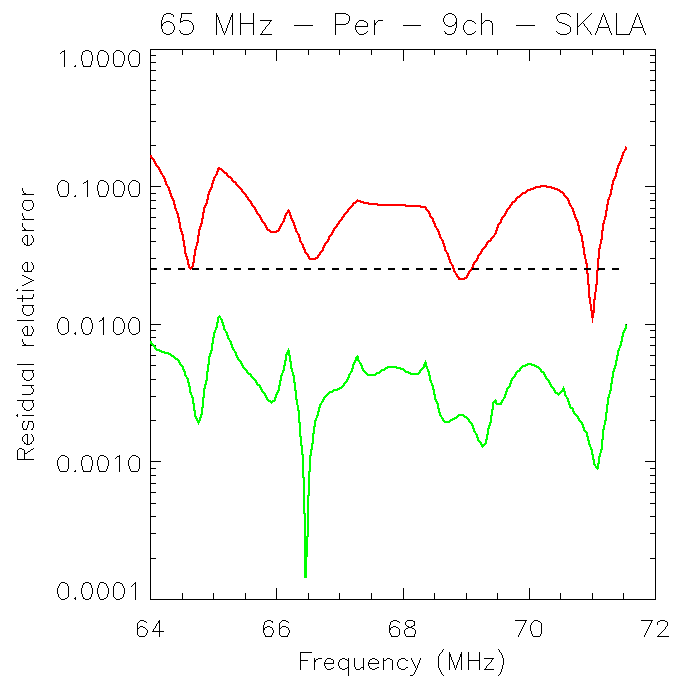}
}
\subfloat[Imperfect 2 calibration model.]{
\includegraphics[width=.45\textwidth]{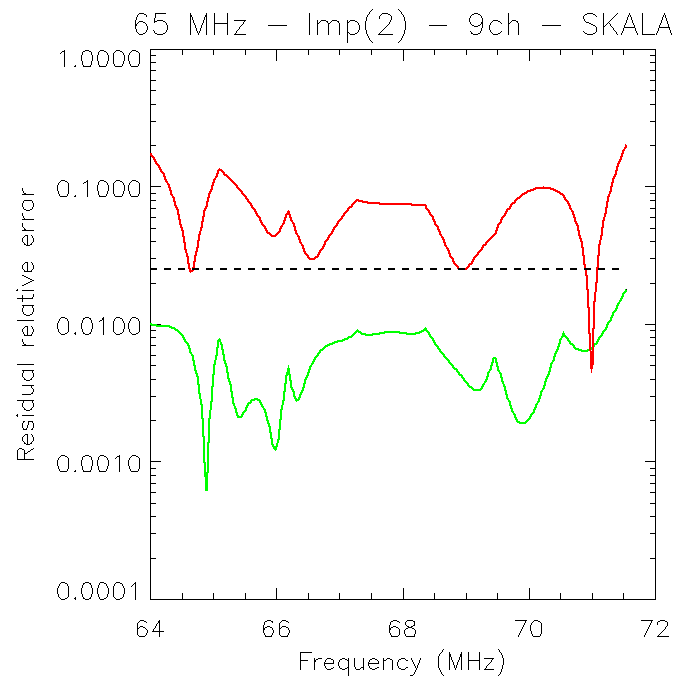}
}
\caption{Relative residual errors for each antenna at 65~MHz, for the perfect and imperfect 2 calibration models and 9 coarse channels of bandwidth. Also plotted is the tolerance level (black, dashed). The wider fitting bandwidth is reflected in the breadth of the frequency axis shown.}
\label{fig:calibration_bw_9}
\end{figure*}

To establish the original specifications, we computed the calibration parameters for each coarse channel, using the central channel and the two contiguous channels. This yields a calibration bandwidth of $\sim$2.5~MHz. In \citet{barry16}, simulations with the MWA suggested that smoothness over a wider bandwidth is also important. Despite this work being for a different experiment and different telescope, here we test the calibration performance of two wider bandwidths: 7 coarse channels ($\sim$5.9~MHz) and 9 coarse channels ($\sim$7.6~MHz). The additional bandwidth will reduce the calibration uncertainty, due to lower noise (more datapoints), but any structure will be preserved, biasing the calibration. We perform these tests close to the low-frequency spectral feature (peak) at $\sim$62--66~MHz. Figure \ref{fig:calibration_bw_7} and \ref{fig:calibration_bw_9} show the performance for perfect and imperfect(2) calibration for 7 and 9 coarse channels, respectively.
The antenna residuals now show different structure, with poorer performance using the SKALA2 antenna. Both antennas also show larger residual errors for fitting 9 channels compared with the narrower bandwidth of Figure \ref{fig:res_sims} and \ref{fig:res_sims_imperfect2}, further suggesting that a third-order polynomial fit is not appropriate over these wider bandwidths.


\subsection*{Polynomial order}
The original tolerances were derived for 2nd-, 3rd-, and 4th-order polynomial fits across the three coarse channels, with (in general), slightly relaxed tolerances for the lower orders. Here we perform the same analysis as described above for 1st-, 2nd- and 4th-order polynomial fits, and compare with the tolerances.

\begin{figure*}
\subfloat[65~MHz.]{
\includegraphics[width=.45\textwidth]{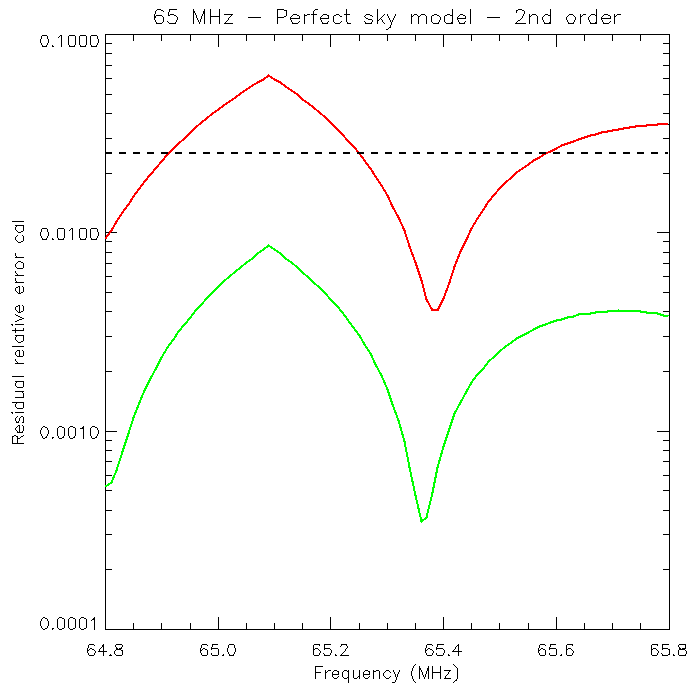}
}
\subfloat[100~MHz.]{
\includegraphics[width=.45\textwidth]{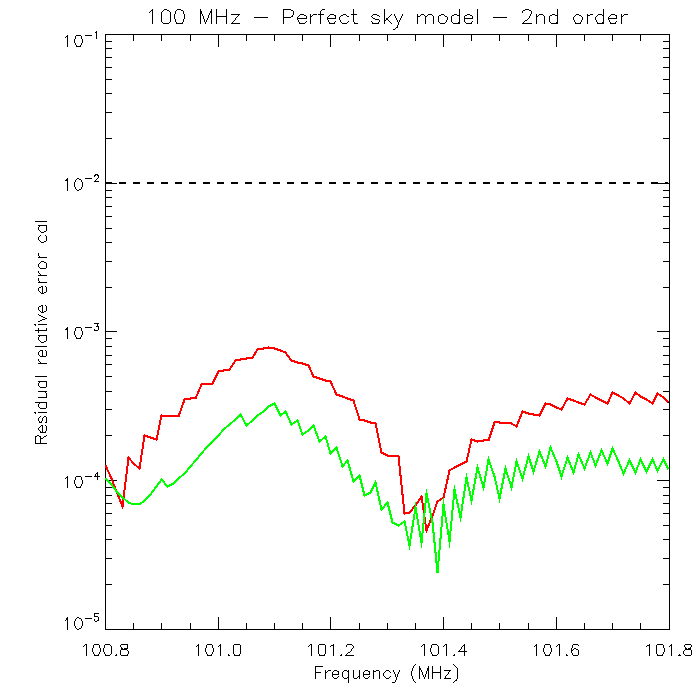}
}
\caption{Residuals in the amplitude of the data, relative to a gain of unity, for two relevant reference frequencies, a perfect calibration model, and a second-order polynomial fit. In all cases, red denotes the SKALA2, and green denotes the SKALA3, design, and black (dashed) shows the tolerance.}
\label{fig:second_order}
\end{figure*}

\begin{figure*}
\subfloat[65~MHz.]{
\includegraphics[width=.45\textwidth]{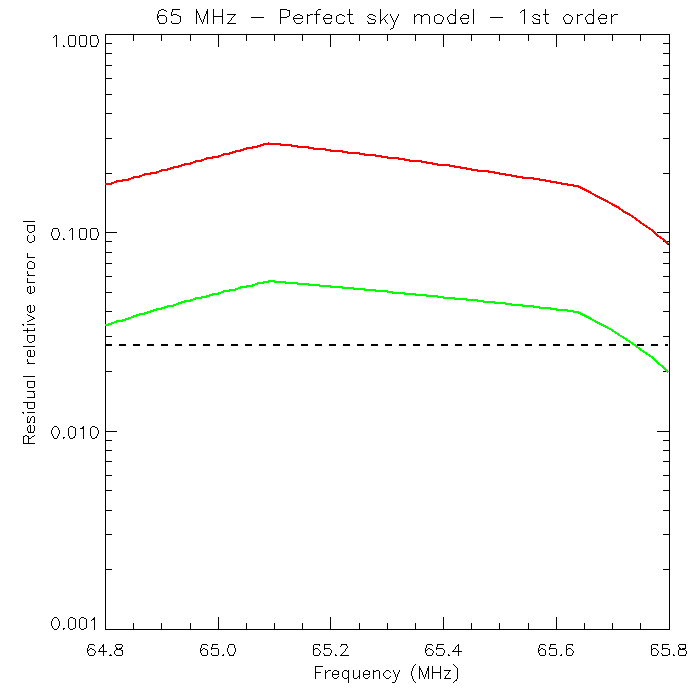}
}
\subfloat[100~MHz.]{
\includegraphics[width=.45\textwidth]{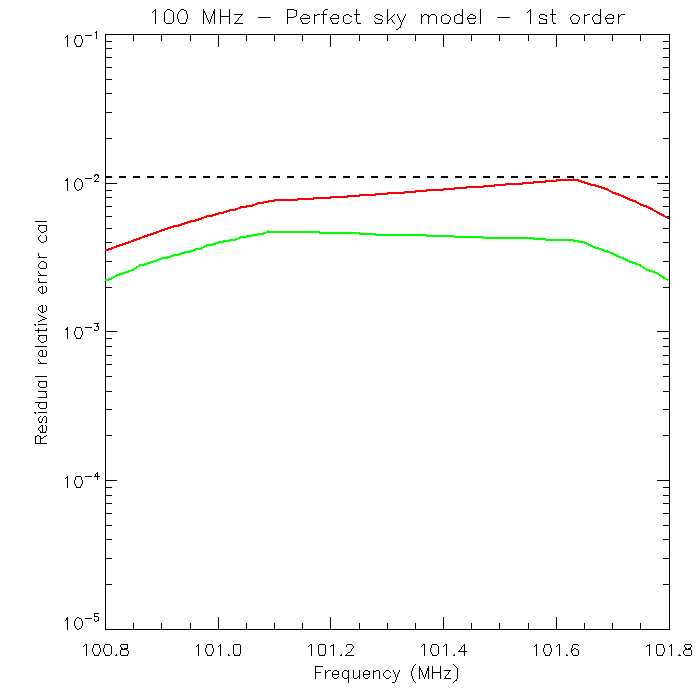}
}
\caption{Residuals in the amplitude of the data, relative to a gain of unity, for two lower relevant reference frequencies, a perfect calibration model, and a first-order polynomial fit. In all cases, red denotes the SKALA2, and green denotes the SKALA3, design, and black (dashed) shows the tolerance.}
\label{fig:first_order}
\end{figure*}

The 4th-order fits showed poorer calibration performance (larger residuals) than the 3rd-order fits shown in the previous section. The 2nd-order fits showed improved performance. Figure \ref{fig:second_order} shows the SKALA2 and SKALA3 results for each of the lower two frequencies and a 2nd-order fit.
Figure \ref{fig:first_order} shows the 1st-order fits.
Comparison of the two lower-order fits demonstrates that the best fitting polynomial may be frequency-dependent, with higher curvature at the low end of the band better represented by a second-order polynomial than a linear fit (first-order).

\begin{figure*}
\subfloat[SKALA2.]{
\includegraphics[width=.45\textwidth]{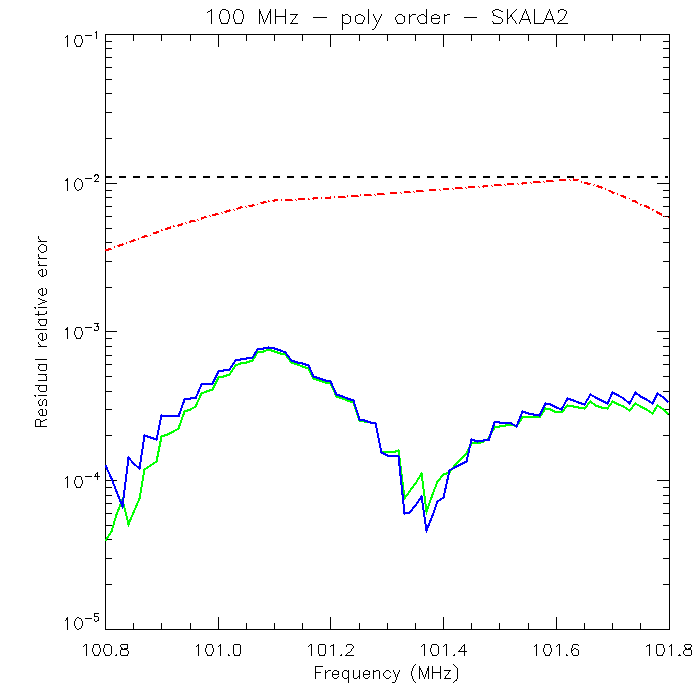}
}
\subfloat[SKALA3.]{
\includegraphics[width=.45\textwidth]{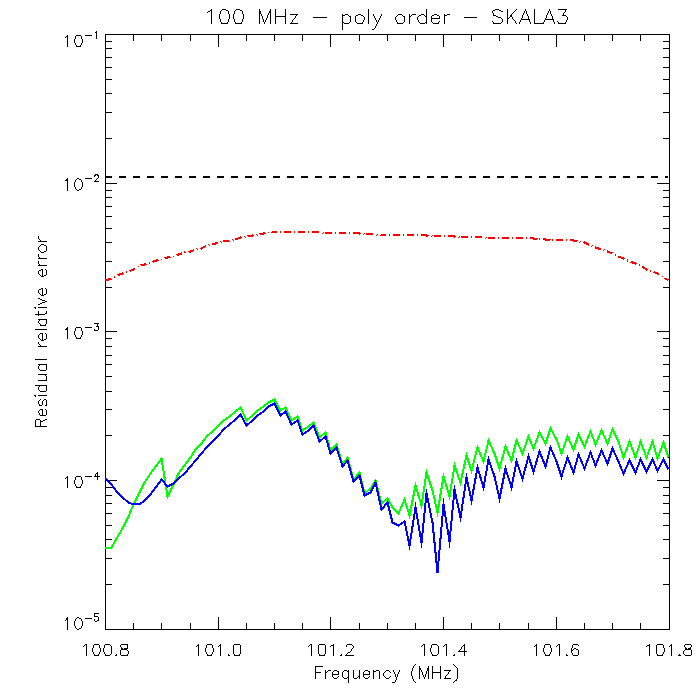}
}
\caption{Relative residual errors for each antenna at 100~MHz, for the a perfect calibration model and varying polynomial orders (green, $n$=3; blue, $n$=2; red dashed, $n$=1). Also plotted is the tolerance level (black, dashed).}
\label{fig:poly_150}
\end{figure*}

In the same vein, all polynomial orders can be overplotted at 100~MHz (Figure \ref{fig:poly_150}).
The antennas show different behaviour with polynomial order, but the general trend is for a better fit for a higher-order polynomial. The tolerance for a second-order polynomial is also shown. This behaviour reflects the fact that, across most of the band, the simulated gain measurements from both antennas is approximately quadratic over a coarse channel. 

\subsection*{EDA performance}
\begin{figure*}
\subfloat[65~MHz with a 3 channel fit.]{
\includegraphics[width=.3\textwidth]{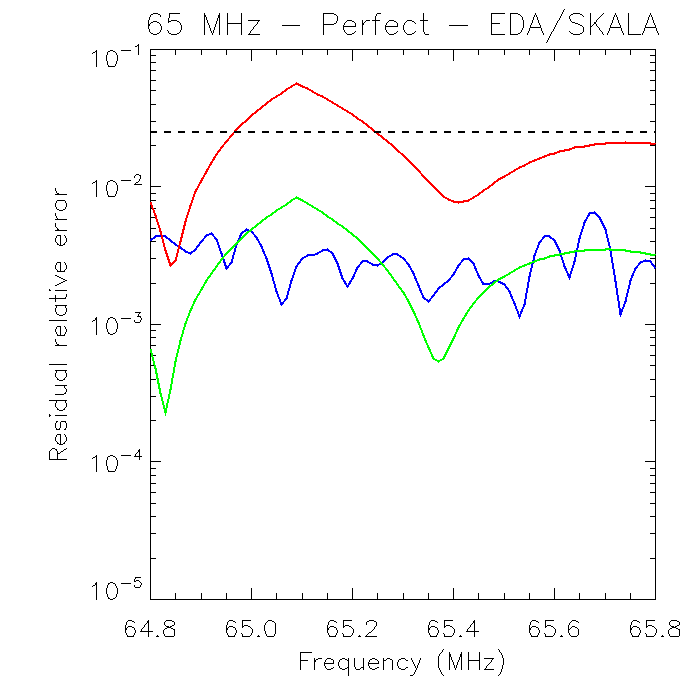}
}
\subfloat[100~MHz with a 3 channel fit.]{
\includegraphics[width=.3\textwidth]{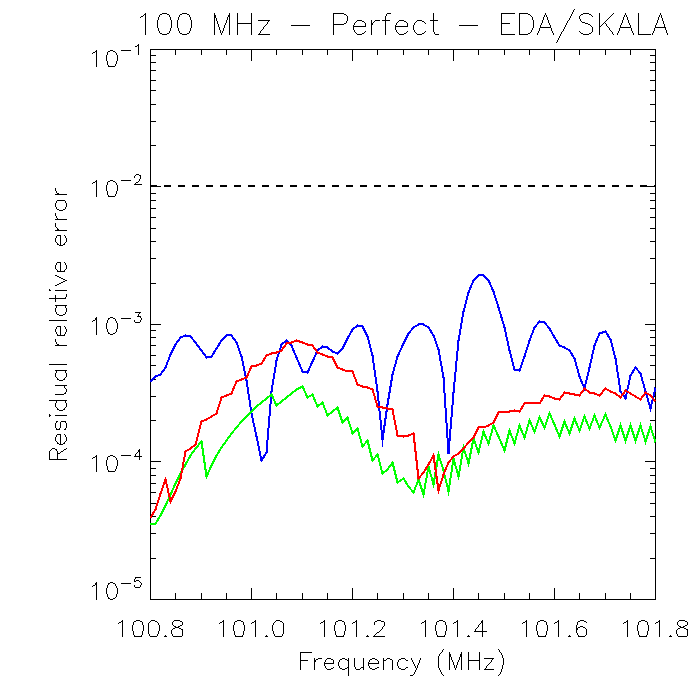}
}
\subfloat[65~MHz with a 7 channel fit.]{
\includegraphics[width=.3\textwidth]{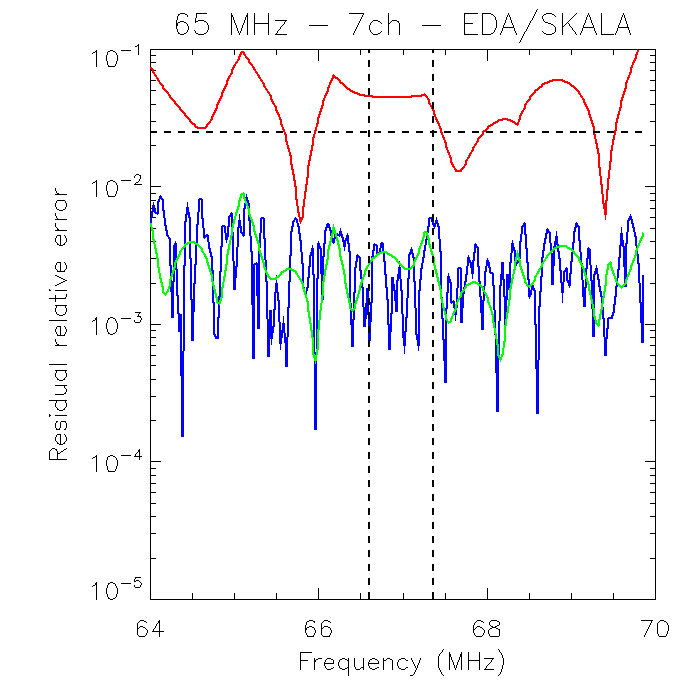}
}
\caption{Residuals in bandpass fits for the SKALA2 (red), SKALA3 (green) and EDA (blue). (Left) 65~MHz with a fit over three coarse channels; (centre) 100~MHz with a fit over three coarse channels; (right) 65~MHz with a fit over seven coarse channels, including two vertical lines to denote the central coarse channel for which that fit is used.}
\label{fig:eda_comparison}
\end{figure*}

We perform identical analysis between the SKALA and EDA antennas for the four same frequencies, three coarse channels, 600-s calibration timescale, and an $n=3$ polynomial fit. This allows a direct comparison of the antenna performance, with the caveats that the EDA complex-valued bandpass measurements rely on real measurements (with low thermal noise). Figure \ref{fig:eda_comparison} shows the EDA and SKALA residuals at 65~MHz (left) and 100~MHz (centre), while a 7-coarse channel fit at 65~MHz is also shown (right).
At the lowest frequency, EDA performance is comparable to SKALA3 at some frequencies, but shows improved performance compared with the SKALA2. Due to the use of actual measurements for the EDA bandpass, there is thermal noise residual structure, not observed in the SKALA residuals. We have attempted to remove this additional residual by smoothing the measured bandpass with a narrow Gaussian kernel (30~kHz; low-pass filter). The general structure observed in the residuals is a function of the stochasticity of the actual measurements with the antenna in the field, above the smoothing kernel scale. At 100~MHz, where the thermal noise plays more of a role in the residual levels, the performance is comparable across all three antennas. The wider band fit similarly yields comparable results, suggestive of the calibration timescale and available sky model being primary drivers for the magnitude of the residuals, rather than actual antenna response spectral structure. In all cases, SKALA3 and the MWA EDA dipole meet the tolerances described in \citet{trottwayth2016}, and show improved performance compared with the SKALA2.

\section{Assessment and conclusions}
Overall, the SKALA3 and MWA antennas perform better than the SKALA2 for instrument calibration using a standard procedure. This is particularly notable below 100~MHz, where the SKALA2 fails to meet the tolerances set out in \citet{trottwayth2016}, even with a perfect calibration model. The associated conclusions point to higher-order polynomials ($n$=2, 3) showing better fits to the simulated gains (lower relative residual errors), but the best order is frequency-dependent. Imperfect calibration models yield reduced performance, but not substantially for the calibration models tested here. A further imperfect model where the calibration only uses sources with apparent flux density $\geq$~1~Jy (not shown in this paper) displays further degraded performance. Deeper calibration and sky models did not yield significantly different results, suggesting that the brightest sky sources are most important for bandpass calibration. Increase in the bandwidth of the fit yielded poorer performance, with the SKALA2 antenna showing poorer performance at the low frequencies associated with a resonant feature in the bandpass, compared with its performance over the smaller bandwidth.

Comparison with in-field measurements from the MWA dipole antenna, as installed in the MWA EDA prototype station, demonstrates that performance gains may be still be available at the lowest frequencies compared with the SKALA2, and comparable performance to SKALA3. 

Important assumptions and caveats:
\begin{itemize}
\item Only four, discrete frequencies were assessed in this work. There is therefore no characterization of other frequencies, where the performance of either antenna model may be substantially degraded;
\item The results are estimated with respect to a core reference antenna, using its spectral shape measurements with respect to all other antennas in the array. Results may differ for a different reference antenna (although, this is not likely);
\item Only polynomial fits have been applied, and no other basis functions are tested;
\item Results assume that the bandpass shapes for each station are independent, and therefore all need to be independently estimated and calibrated;
\item Results assume that calibration solutions are independent between each 600-second calibration cycle, and prior calibration fits are not used and not applicable.
\end{itemize}

The results presented here for SKALA3 show adequate performance to undertake the challenging EoR/CD experiments proposed for SKA1-Low, compared with the SKALA2. SKALA3 is now going to be tested with field measurements under identical environmental and signal chain conditions that will be met by the full array at the MRO, and as part of a set of antennas forming a single SKA station. Use of the measured response of the MWA EDA dipole in calibration performance assessment also demonstrates its adequacy to perform the suite of EoR/CD experiments.

\section*{Acknowledgements}
The authors thank Leon Koopmans and the SKA Office for useful discussions, Budi Juswardy for assistance with the EDA antenna measurements, and the Low Frequency Aperture Array group of SKA for support in development of the SKALA3 antenna.
This research was supported under the Australian Research Council's Discovery Early Career Researcher funding scheme (project number DE140100316), and the  Centre for All-sky Astrophysics (an Australian Research Council Centre of Excellence funded by grant CE110001020). This work was supported by resources provided by the Pawsey Supercomputing Centre with funding from the Australian Government and the Government of Western Australia. We acknowledge the iVEC Petabyte Data Store, the Initiative in Innovative Computing and the CUDA Center for Excellence sponsored by NVIDIA at Harvard University, and the International Centre for Radio Astronomy Research (ICRAR), a Joint Venture of Curtin University and The University of Western Australia, funded by the Western Australian State government. This research was also supported by the Science \& Technology Facilities Council (UK) grant: \textit{SKA, ST/M001393/1} and the University of Cambridge, UK.

\bibliographystyle{mnras}
\bibliography{pubs}

\bsp	
\label{lastpage}
\end{document}